\documentclass[11pt,a4paper]{article}
\usepackage{jheppub}
\usepackage[utf8]{inputenc}
\usepackage{latexsym, graphicx} 
\usepackage{amsmath,amsfonts,amssymb}
\usepackage{mathrsfs}
\usepackage{tensor}

\newcommand*{\scri}{\ensuremath{\mathscr{I}}}

\newcommand*{\dd}{\mathop{}\!d}

\newcommand{\SCD}{S_{\text{CD}}}
\newcommand{\Cp}{C_{\text{plane}}}

\title{Celestial IR divergences and the effective action\\ of supertranslation modes}
\author[a]{Kevin Nguyen,}
\author[b]{Jakob Salzer}
\emailAdd{kevin.nguyen@kcl.ac.uk}
\emailAdd{jsalzer@fas.harvard.edu}
\affiliation[a]{Department of Mathematics, King's College London, London, United Kingdom}
\affiliation[b]{Center for the Fundamental Laws of Nature, Harvard University, Cambridge, MA02138, USA}

\abstract{Infrared divergences in perturbative gravitational scattering amplitudes have been  recently argued to be governed by the two-point function of the supertranslation Goldstone mode on the celestial sphere. We show that the form of this celestial two-point function simply derives from an effective action that also controls infrared divergences in the symplectic structure of General Relativity with asymptotically flat boundary conditions. This effective action finds its natural place in a path integral formulation of a celestial conformal field theory, as we illustrate by re-deriving the infrared soft factors in terms of celestial correlators. Our analysis relies on a well-posed action principle close to spatial infinity introduced by Comp\`ere and Dehouck.}

\begin{document}

\maketitle
\flushbottom

\section{Introduction}
\label{sec:intro}

Scattering amplitudes in perturbative quantum gravity infamously suffer from infrared (IR) divergences as a result of the real emission and virtual exchange of soft gravitons. Building upon earlier work performed in the context of quantum electrodynamics, Weinberg however showed that these IR divergences cancel in measurable cross-sections for which an arbitrary number of soft gravitons contribute \cite{Weinberg:1965nx}. Such cancellations are made possible thanks to the \textit{factorization} of any scattering amplitude $\mathcal{A}$ into soft and hard parts,
\begin{equation}
\label{eq:factorization}
\mathcal{A}=\mathcal{A}_{\text{soft}}\,  \mathcal{A}_{\text{hard}}\,,
\end{equation}
where 
$\mathcal{A}_{\text{soft}}$ bears all dependencies on the soft quanta and IR divergences. In the case of a soft emission in particular, $\mathcal{A}_{\text{soft}}$ is determined by Weinberg's soft graviton theorem \cite{Weinberg:1965nx}. On the other hand and of direct interest to the present work, the soft factor resulting from the exchange of virtual soft gravitons between $n$ external hard massless particles takes the simple form \cite{Weinberg:1965nx,Naculich:2011ry,White:2011yy,Himwich:2020rro,Arkani-Hamed:2020gyp}
\begin{equation}
\label{eq:Asoft}
\mathcal{A}_{\text{soft}}=\exp\left[-\frac{2 G \log \Lambda_0}{\pi} \sum_{i,j=1}^n \eta_i \eta_j \omega_i \omega_j\,  |x_i-x_j|^2 \ln |x_i-x_j|^2 \right]\,,
\end{equation}
where $G$ is Newton's constant, and each massless hard particle is characterized by its frequency $\omega_i>0$ and a stereographic coordinate $x_i \in \mathcal{S}$ on the celestial sphere at which it enters or exits the process depending on whether it is ingoing ($\eta_i=-1$) or outgoing ($\eta_i=+1$), respectively.
Weinberg introduced the parameter $\Lambda_0=E_+/E_->1$ as the ratio of two energy cutoffs. The first arbitrarily separates between soft and hard gravitons. In practice, it is fixed by the detector sensitivity such that soft gravitons are those missed by the detector. The second energy scale $E_-$ is an IR regulator that must eventually be sent to zero.\footnote{Within dimensional regularization, $\log \Lambda_0$ is replaced by $1/\epsilon$ with $d=4-\epsilon$ \cite{Naculich:2011ry}.} 

More recently, connections between IR divergences and the asymptotic structure of asymptotically flat spacetimes in General Relativity have been uncovered. In particular, conservation laws associated with the group of asymptotic symmetries discovered long ago by Bondi, van der Burg, Metzner and Sachs (BMS) \cite{Bondi:1962px,Sachs:1962wk} have been identified with the soft graviton theorems \cite{Strominger:2013jfa,Strominger:2014pwa,He:2014laa}. Furthermore, it has been argued that the soft factor \eqref{eq:Asoft} resulting from the exchange of virtual soft gravitons can be written in terms of correlation functions of the Goldstone mode $C(x)$ of spontaneously broken {supertranslations} that form a subset of the BMS symmetries~\cite{Himwich:2020rro},
\begin{equation}
\label{eq:A soft}
\mathcal{A}_{\text{soft}}=\exp\left[-\frac{1}{2} \sum_{i\neq j}^n \eta_i \eta_j \omega_i \omega_j \langle C(x_i) C(x_j) \rangle \right]\,.
\end{equation}
By comparison with \eqref{eq:Asoft}, the authors of \cite{Himwich:2020rro} inferred the form of the celestial two-point correlation function of the supertranslation Goldstone mode,
\begin{equation}
\label{eq:C correlator}
\langle C(x) C(y) \rangle=\frac{4G}{\pi} \log \Lambda_0 \, |x-y|^2 \log |x-y|^2\,.
\end{equation}

The goal of the present work is to provide an independent derivation of the celestial correlator \eqref{eq:C correlator}, which in particular will not rely at any point on the evaluation of scattering amplitudes. Rather, our derivation will be based on the behavior of the gravitational field at spatial infinity together with a careful analysis of the corresponding action principle. In particular, we identify a component of the gravitational action which at the same time governs IR divergences on the celestial sphere and controls the supertranslation Goldstone mode in a way that yields the celestial two-point function~\eqref{eq:C correlator}. We call it the \textit{infrared effective action}.

Our derivation of a two-dimensional effective action for supertranslation Goldstone modes can be regarded as part of the broader celestial conformal field theory (CFT) program. This ongoing effort finds its origins in the observation that the original BMS group can be enlarged to include two copies of the Virasoro symmetry \cite{Barnich:2010eb}. This suggests that gravitational scattering amplitudes can be recast in the form of correlation functions of a two-dimensional CFT defined on the celestial sphere. The celestial CFT program aims both at making this correspondence precise and at describing the properties of this putative celestial CFT \cite{deBoer:2003vf,Barnich:2010eb,He:2015zea,Kapec:2016jld,Pasterski:2016qvg,He:2017fsb,Pasterski:2017ylz,Donnay:2018neh,Himwich:2019dug,Pate:2019mfs,Ball:2019atb,Pate:2019lpp,Donnay:2020guq,Puhm:2019zbl,Himwich:2020rro,Arkani-Hamed:2020gyp,Atanasov:2021oyu,Guevara:2021abz,Stieberger:2018edy,Stieberger:2018onx,Fan:2019emx,Fan:2020xjj,Fan:2021isc,Fotopoulos:2019tpe,Fotopoulos:2019vac,Fotopoulos:2020bqj,Cheung:2016iub,Guevara:2019ypd,Banerjee:2019aoy,Banerjee:2019tam,Adamo:2019ipt,Nandan:2019jas,Pasterski:2017kqt,Schreiber:2017jsr,Ebert:2020nqf,Banerjee:2020kaa,Law:2019glh,Law:2020xcf,Atanasov:2021cje,Crawley:2021ivb,Pasterski:2021fjn,Kalyanapuram:2020epb,Kalyanapuram:2021bvf,Pasterski:2021dqe}. Indeed, this reformulation of the gravitational S-matrix is often viewed as a potential stepping stone towards the uncovery of a holographic duality in asymptotically flat spacetimes, in a way analogous to the celebrated AdS/CFT correspondence \cite{Maldacena:1997re,Gubser:1998bc,Witten:1998qj}. We make progress in this direction by showing that the newly found infrared effective action finds a natural place in a path integral formulation of the celestial CFT.   

An important open issue in this context is a better understanding of the IR sector of asymptotically flat gravity and its description in terms of a celestial CFT. Supertranslations and superrotations are spontaneously broken, leading to an infinite degeneracy of gravitational vacua \cite{Ashtekar:1981hw,Strominger:2013jfa}; see \cite{Compere:2016jwb,Compere:2018ylh,Adjei:2019tuj} for an explicit construction of the corresponding metric solutions. Here, we turn our attention to the supertranslation Goldstone mode $C(x)$ and its conjectured celestial two-point correlation function \eqref{eq:C correlator}. A discussion of the superrotation Goldstone modes and their role in the celestial CFT can be found in \cite{Nguyen:2020hot,Nguyen:2021dpa}.

Most of the previous discussion has implicitly referred to the structure of asymptotically flat spacetimes at null infinity $\scri$, where asymptotic states of massless particles and radiation fields can be defined. In particular, the BMS group and its extensions are purely determined through a local analysis of the gravitational field near $\scri$. The story to be developed in this paper however relies on the study of the gravitational field at spatial infinity~$i^0$, which has received significantly less attention than its null cousin. The study of asymptotic symmetries at $i^0$ similarly reveals an infinite-dimensional enhancement of the Poincar\'e group, containing in particular \textit{supertranslations at spatial infinity (spi)} that are naturally identified as counterpart of BMS supertranslations; see \cite{Compere:2011ve,Troessaert:2017jcm, Henneaux:2018cst,Henneaux:2018hdj} and references therein. The precise correspondence between asymptotic structures at null and spatial infinity has however not been fully worked out. Recent progress in matching of symmetries and charges from $i^0$ to $\scri$ includes \cite{Troessaert:2017jcm,Prabhu:2019fsp,Prabhu:2019daz}.

In this paper, we derive the celestial two-point function \eqref{eq:C correlator} of the supertranslation Goldstone mode by analyzing the gravitational field equations and action principle near spatial infinity $i^0$. This allows us to single out a component of the action governing celestial IR divergences together with the supertranslation Goldstone mode.
We start in section~\ref{sec:inought} with a review of the Beig--Schmidt formulation of General Relativity close to spatial infinity $i^0$ \cite{Beig:1982bs,Beig}. We adopt the action principle of Comp\`ere and Dehouck which was shown to yield well-defined variations together with a finite symplectic structure \cite{Compere:2011ve}. This action principle comes with falloff and boundary conditions that admit spi-supertranslations as part of the asymptotic symmetry group, and we identify the corresponding Goldstone mode.
Following earlier work by Troessaert \cite{Troessaert:2017jcm}, in section~\ref{sec:scrimatch} we make the connection with familiar quantities at null infinity $\scri$. In particular, we provide the map between supertranslation Goldstone modes at $i^0$ and $\scri$. 
In section~\ref{sec:effective action}, we derive the celestial two-point function \eqref{eq:C correlator} from the component of the action which governs celestial IR divergences. The latter can be viewed as an \textit{effective action} for the supertranslation Goldstone mode in the IR-divergent theory. In section~\ref{sec:path integral}, we show that this infrared effective action finds a natural place in the evaluation of the soft factor \eqref{eq:A soft} as the expectation value of vertex operators in the celestial CFT \cite{Himwich:2020rro}. More specifically, we show that it is the appropriate action to consider in a path integral representation of this celestial correlator. We end with a discussion of the results and possible future directions. 

\paragraph{Conventions.} We denote four-dimensional spacetime indices with greek letters $\alpha, \beta,...$, and use $\nabla_\alpha$ as the corresponding covariant derivative. We denote three-dimensional indices with roman letters $a,b,c,...$, and use $(h_{ab}\,, D_a)$ for the metric and covariant derivative associated with three-dimensional de Sitter spacetime. Capital roman indices $A,B,...$ refer to arbitrary coordinates on the (celestial) sphere~$\mathcal{S}$ equipped with $(\gamma_{AB}\,,D_A)$, while indices $i,j,...$ specifically refer to stereographic coordinates, i.e., cartesian coordinates on the euclidean plane. Everything should be rather obvious in context.

\section{Gravity at spatial infinity}
\label{sec:inought}
We start by recalling the formulation of General Relativity near spatial infinity $i^0$, putting emphasis on the variational principle, the required boundary conditions and the resulting set of asymptotic symmetries. 

To properly describe spatial infinity, we write the metric in Beig--Schmidt form~\cite{Beig:1982bs} 
\begin{equation}
  \label{eq:BS}
  \dd s^2=N^2\dd\rho^2+H_{ab}\left(N^a \dd \rho+ \dd x^a\right) (N^b \dd \rho+ \dd x^b)\,,
\end{equation}
where
\begin{subequations}
\label{eq:metric components}
\begin{align}
N&=1+\frac{\sigma}{\rho}\,,\\
H_{ab}N^{b}&=o(\rho^{-1})\,,\\
H_{ab}&=\rho^2\left(h_{ab}+\rho^{-1}h^{(1)}_{ab}+\frac{\log \rho}{\rho^2}\, i_{ab}+\rho^{-2}h^{(2)}_{ab}+o(\rho^{-2})\right)\,.
\end{align}
\end{subequations}
Spatial infinity is approached as $\rho \to \infty$, and \eqref{eq:metric components} should be understood as the corresponding asymptotic expansion of the metric. 
As part of the boundary conditions, we do not allow $h_{ab}$ to fluctuate and for consistency with Einstein's equations, we further fix it to coincide with the metric on the unit three-dimensional hyperboloid $\mathcal{H}$ of constant positive curvature (also known as de Sitter spacetime). Hence, we require it to obey
\begin{equation}
\label{eq:R condition}
R\left[h\right]_{ab}=2h_{ab}\,.
\end{equation}
It is in fact convenient to treat $h_{ab}$ as a genuine metric on $\mathcal{H}$, and define the corresponding covariant derivative $D_a$. All three-dimensional indices $a,b,c,..$ are raised and lowered with this metric. The logarithmic term proportional to $i_{ab}$ is necessary and sufficient in order to describe solutions to Einstein's equations when $\sigma$ or $h^{(1)}_{ab}$ do not obey parity conditions \cite{Beig,Compere:2011ve}, as we presently assume. We will comment further on parity conditions in section~\ref{sec:scrimatch}.

In the following it will be convenient to trade the subleading component $h^{(1)}_{ab}$ for the combination
\begin{equation}
\label{eq:defk}
k_{ab}\equiv h^{(1)}_{ab}+2\sigma h_{ab}\,.
\end{equation}
The physical meaning of the asymptotic fields $\sigma$ and $k_{ab}$ can be deduced by noting that the quantities
\begin{equation}
\label{eq:Weyl}
E^{(1)}_{ab}=-\left(D_aD_b+h_{ab}\right)\sigma\,, \qquad B^{(1)}_{ab}=\frac{1}{2}\epsilon\indices{_a^{cd}}D_ck_{db}\,,
\end{equation}
are the leading (non-vanishing) terms of the electric and magnetic parts of the Weyl tensor, respectively. The electric part of the Weyl tensor relates to the mass and momentum aspects of the gravitational field, while its magnetic part carries information about gravitational radiation \cite{Ashtekar:1991vb}. We see that the fields $\sigma$ and $k_{ab}$ play the role of potentials for these two fields.

\begin{figure}
    \centering
    \includegraphics[scale=0.7]{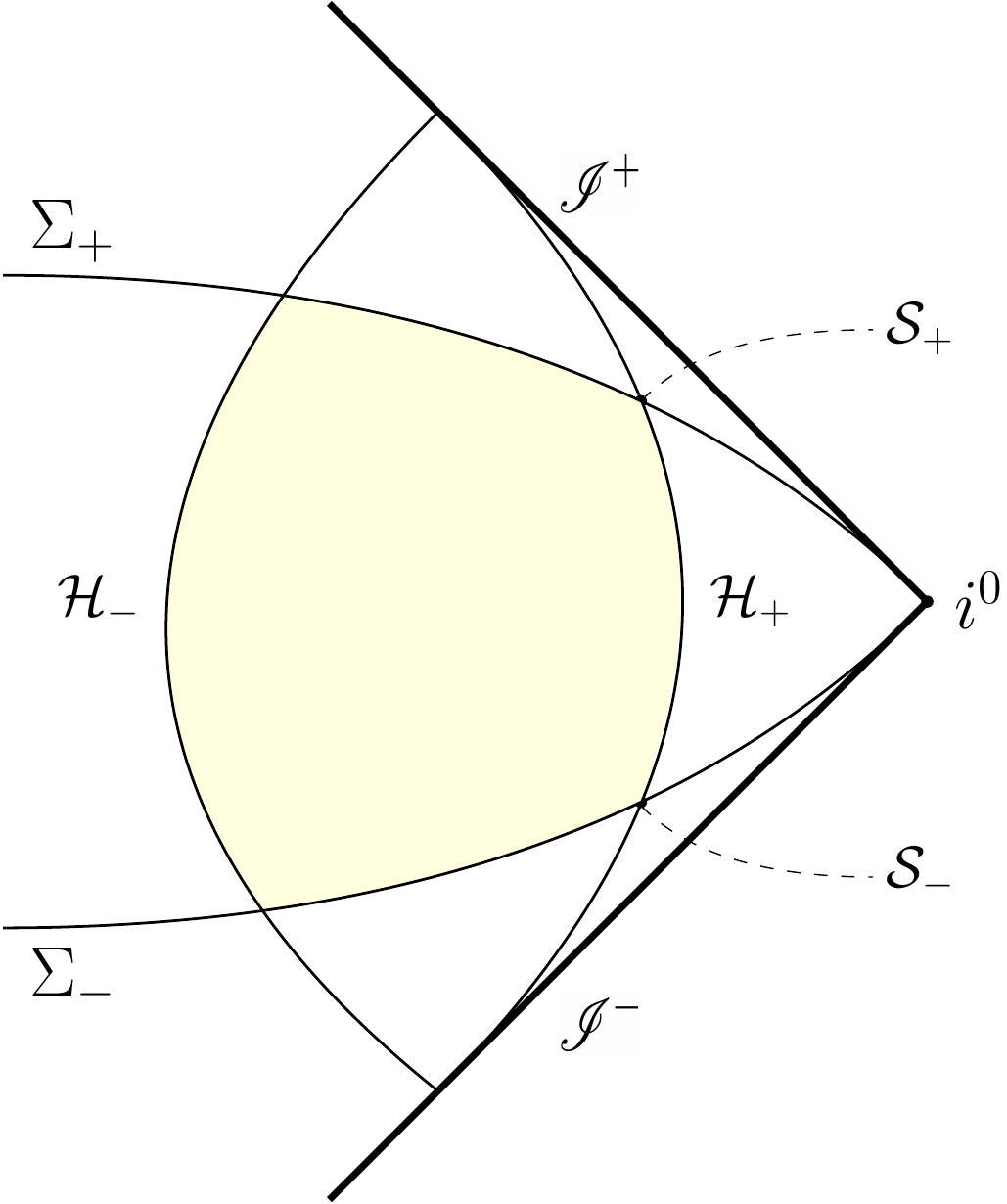}
    \caption{Schematic representation of the geometric setup required to define the action principle near spatial infinity $i^0$. The de Sitter hyperboloids $\mathcal{H}_\pm$ are surfaces of constant $\rho=\Lambda_\pm$. The action principle is defined for the yellow region of spacetime bounded by $\mathcal{H}_\pm$ and two Cauchy surfaces $\Sigma_\pm$. The intersections of $\mathcal{H}_+$ with $\Sigma_\pm$ are denoted $\mathcal{S}_\pm$ and possess the geometry of a two-sphere. In the limit $\Lambda_+ \to \infty$, these two-spheres are identified with $\scri^+_-$ and $\scri^-_+$, respectively.}
    \label{fig:penrose}
\end{figure}

We introduce two cutoff surfaces: one at $\rho=\Lambda_-$ below which the asymptotic expansion \eqref{eq:metric components} is not a valid approximation to the metric, and one at $\rho=\Lambda_+>\Lambda_-$ where boundary conditions will be imposed. These surfaces, respectively denoted $\mathcal{H}_+$ and $\mathcal{H}_-$, have the geometry of a de Sitter hyperboloid; see figure~\ref{fig:penrose}. In this paper, we will consider the behavior of the gravitational field in the region $\Lambda_- < \rho < \Lambda_+ $ bounded by these two cutoff surfaces. We also work in the regime $\Lambda_- \gg 1$ in Planck units, and at the end of the day we remove the infrared cutoff by taking the limit $\Lambda_+ \to \infty$.
These two parameters can be regarded as the respective inverses of the energy cutoffs in Weinberg's factorization formula \eqref{eq:Asoft}, i.e., $\Lambda_+\sim 1/E_-$ and $\Lambda_-\sim 1/E_+$. This point will become manifest in section~\ref{sec:path integral} when re-deriving the soft factor \eqref{eq:A soft}. 
We also note in passing that the above asymptotic expansion implies that the fields $\sigma\,, h_{ab}\,,h^{(1)}_{ab}$ have mass dimension one, zero, and minus one, respectively, while the coordinates $x^a$ are dimensionless.

\paragraph{Action principle.} In order to have a well-posed action principle, one needs to impose additional boundary conditions on the phase-space described by \eqref{eq:BS}-\eqref{eq:metric components}. We adopt those of Comp\`ere and Dehouck \cite{Compere:2011ve} which are less restrictive than all other boundary conditions previously considered in the literature, 
\begin{equation}
\label{eq:k conditions}
k\equiv h^{ab}k_{ab}=0, \qquad D^a k_{ab}=0\,.
\end{equation}
On the resulting restricted phase-space, a suitable action to consider is \cite{Compere:2011ve,Virmani:2011gh}
\begin{equation}
  \label{eq:totaction}
S=S_{\text{EH}}+S_{\mathcal{H}_+}\,,
\end{equation}
where $S_{\text{EH}}$ denotes the standard Einstein--Hilbert action
\begin{equation}
\label{eq:EH}
S_{\text{EH}} =\frac{1}{16\pi G}\int_{\mathcal{M}}\dd^4x\, \sqrt{-g}\, R\,,
\end{equation}
and $S_{\mathcal{H}_+}$ is the sum of the Gibbons--Hawking--York (GHY) and Mann--Marolf (MM) countertems required to make the variational principle well-posed at the spatial boundaries,
\begin{equation}
\label{eq:hypaction}
S_{\mathcal{H}_+}=\frac{1}{8\pi G}\int_{\mathcal{H}_+}\dd^3x\, \sqrt{-H}\, (K-\hat{K})\,.
\end{equation}
Here, $K$ is the extrinsic curvature while $\hat{K}=H^{ab}\hat{K}_{ab}$ is an intrinsic boundary quantity defined implicitly\footnote{Although we will not make use of them, explicit formulas for $\hat{K}_{ab}$ are also available \cite{Visser:2008gx}.} by the relation \cite{Mann:2005yr}
\begin{equation}
  \label{eq:khatdef}
  R_{ab}[H]=\hat{K}_{ab}\hat{K}-\hat{K}\indices{_a^c}\hat{K}_{cb}\,,
\end{equation}
where $R_{ab}[H]$ is the Ricci tensor constructed from the induced metric $H_{ab}$. Keeping the leading term $h_{ab}$ fixed as previously discussed, it has indeed been shown that the boundary contribution to the action variation is  \cite{Compere:2011ve,Virmani:2011gh}
\begin{equation}
\label{eq:kvariation}
\delta S|_{\mathcal{H}_+}= \frac{1}{16\pi G}\int_{\mathcal{H}_+}\dd^3x\, \sqrt{-h} \left(D^{a}D^{b}\sigma+h^{ab}\sigma\right)\delta k_{ab}\,.
\end{equation}
Integrating by parts, one finds that this contribution vanishes thanks to the boundary conditions \eqref{eq:k conditions}. However, this is not yet enough to ensure well-posedness of the action principle. Indeed, Comp\`ere and Dehouck also found linearly and logarithmically divergent contribution at the past and future boundaries $\mathcal{S}_\pm$ of $\mathcal{H}_+$,
\begin{align}
\label{eq:delta S log}
\delta S|_{\mathcal{S}_\pm}=&\pm \frac{\log \Lambda_+}{16\pi G} \int_{\mathcal{S}_\pm} \dd S_a \left(4 \delta \sigma D^a \sigma+\frac{1}{2} \delta k_{bc}\, D^a k^{bc}-\delta k_{bc}\, D^c k^{ab} \right)\\
\nonumber
&+\delta \left(\Lambda_+\, \mathcal{R}_{\mathcal{S}_\pm}+\log \Lambda_+\, \mathcal{R}^{(log)}_{\mathcal{S}_\pm}+\log^2 \Lambda_+\, \mathcal{R}^{(log\, 2)}_{\mathcal{S}_\pm} \right)+O(\Lambda_+^0)\,,
\end{align}
where $\dd S_a$ is the future-directed surface-element on $\mathcal{S}_\pm$. 
The details of this computation and expressions for $\mathcal{R}_{\mathcal{S}_\pm}\,, \mathcal{R}^{(log)}_{\mathcal{S}_\pm}$ and $\mathcal{R}^{(log\, 2)}_{\mathcal{S}_\pm}$ are given in appendix~\ref{app:variational principle}. The total variation in the second line can be cancelled by subtracting these terms from the original action.
Cancellation of the logarithmically divergent term in the first line of \eqref{eq:delta S log} requires the introduction of yet another boundary counterterm localized on the hyperboloid~$\mathcal{H}_+$,
\begin{equation}
\label{eq:S extra}
\SCD=\frac{\log \Lambda_+}{4\pi G}\left( S^{(\sigma)}+S^{(k)}\right)\,,
\end{equation}
where
\begin{subequations}
\begin{align}
S^{(\sigma)}&=-\frac{1}{2}\int_{\mathcal{H}} \dd^3x\, \sqrt{h}\, \left(D^a \sigma D_a \sigma-3 \sigma^2 \right)\,,\\
S^{(k)}&=-\frac{1}{8}\int_{\mathcal{H}} \dd^3x\, \sqrt{h}\, \left(\frac{1}{2} D_a k_{bc}\, D^a k^{bc}-D_a k_{bc} D^b k^{ac}-\frac{3}{2} k^{ab} k_{ab} \right)\nonumber\\
&=\frac{1}{16}\int_{\mathcal{H}} \dd^3x\, \sqrt{h}\, \left[k^{ab} D^2 k_{ab} -3k^{ab}k_{ab}+D^c\left(2k^{ab}D_b k_{ac}-k^{ab}D_c k_{ab} \right) \right]\,.
\end{align}
\end{subequations}
Although we will not make explicit use of this observation, we note that the action $S^{(k)}$ is the action for a partially massless graviton in dS$_3$ \cite{Deser:1983mm}.
The equations of motion resulting from these boundary actions,
\begin{equation}
\label{eq:eom sigma and k}
(D^2+3)\sigma=0\,, \qquad (D^2-3)k_{ab}=0\,,
\end{equation}
already follow from expanding Einstein's equations in Beig--Schmidt gauge \eqref{eq:BS}. The variation of $\SCD$ therefore only produces a term localized at $\mathcal{S}_\pm$ that precisely cancels against \eqref{eq:delta S log}. Hence, the total action we consider is
\begin{equation}
\label{eq:S total}
S_{\text{total}}=S-\Lambda_+\, \Delta \mathcal{R}_{\mathcal{S}}-\log \Lambda_+\, \Delta \mathcal{R}^{(log)}_{\mathcal{S}}-\log^2 \Lambda_+\, \Delta \mathcal{R}^{(log\, 2)}_{\mathcal{S}}+\SCD\,,
\end{equation}
where $\Delta \mathcal{R}_{\mathcal{S}} \equiv \mathcal{R}_{\mathcal{S}_+}-\mathcal{R}_{\mathcal{S}_-}$ and similarly for the other terms.
In deriving this action, we restricted our attention to the outer spatial boundary $\mathcal{H}_+$. The exact same reasoning also applies to the inner boundary $\mathcal{H}_-$, which can be accounted for by making the replacements $\Lambda_+ \mapsto \Lambda_+-\Lambda_-$ and $\log \Lambda_+ \mapsto \log \Lambda_+/\Lambda_-=\log \Lambda$ in \eqref{eq:S extra} and \eqref{eq:S total}.

As we demonstrate in appendix~\ref{sec:GHY and MM}, the term $\Delta \mathcal{R}_{\mathcal{S}}$ is proportional to the difference in the total mass between the slices $\Sigma_+$ and $\Sigma_-$ and thus vanishes when the equations of motion hold. Similarly, $\mathcal{R}^{(log\, 2)}_{\mathcal{S}}$ vanishes when the constraints on $i_{ab}$ implied by Einstein's equations hold. This is not the case for the logarithmic contribution $\mathcal{R}^{(log)}_{\mathcal{S}}$ though, and one finds that the additional term $\SCD$ is needed to cancel this divergent contribution. As Comp\`ere and Dehouck showed, the inclusion of $\SCD$ also renders the total symplectic structure finite, and in particular yields conserved and finite expressions for Lorentz and spi-supertranslation charges.

\paragraph{Supertranslations and Goldstone modes.} Asymptotic symmetries are built from diffeomorphisms that preserve the gauge \eqref{eq:BS}, falloffs \eqref{eq:metric components} and boundary conditions \eqref{eq:R condition} and \eqref{eq:k conditions}, and that are in addition associated with nonzero charges. In contrast, diffeomorphisms associated with vanishing charges are considered proper gauge redundancies.
As described in \cite{Compere:2011ve} and earlier works referenced therein, in the present context the group of asymptotic symmetries contains Poincar\'e transformations, \emph{(spi) supertranslations} and \emph{logarithmic translations}. For the purpose of the present work, we will restrict our attention to supertranslations, parametrized by the function $\omega(x^a)$ through the coordinate transformation
\begin{equation}
\rho \mapsto \rho+\omega+o(\rho^0)\,, \qquad  x^a\mapsto x^a+\rho^{-1}D^a\omega+o(\rho^{-1})\,.
\end{equation}
Supertranslations act on the various asymptotic fields as
\begin{equation}
  \label{eq:STalpha}
  \delta_\omega \sigma=\delta_\omega h_{ab}=0,\qquad \delta_{\omega}k_{ab}=2(D_aD_b+h_{ab})\omega\,.
\end{equation}
Due to the boundary condition \eqref{eq:k conditions}, the function $\omega(x^a)$ must satisfy
\begin{equation}
  \label{eq:STmode}
\left(D^2+3\right) \omega=0\,.
\end{equation}
The charges associated to these supertranslations at spatial infinity were computed in \cite{Compere:2011ve} and read
\begin{align}
  \label{eq:STcharges}
  Q[\omega]=\frac{1}{4\pi G}\int_{\mathcal{S}} \dd S^a \left(D_a\sigma \,\omega- D_a\omega \,\sigma\right)\,,
\end{align}
where $\mathcal{S}$ can be any spacelike cut of the hyperboloid $\mathcal{H}$. Using the above properties, it is straightforward to show that these charges are conserved and thus independent of the cut.

For stationary spacetimes, $B^{(1)}_{ab}$ defined in \eqref{eq:Weyl} vanishes and this absence of radiation directly translates to
\begin{equation}
  \label{eq:norad}
  D_{[a}k_{b]c}=0\,.
\end{equation}
On the three-dimensional hyperboloid $\mathcal{H}$, a symmetric tensor field $k_{ab}$ obeying \eqref{eq:norad} can be written in terms of a scalar potential \cite{Ashtekar:1978zz,Ashtekar:1984zz,Prabhu:2019daz},
\begin{equation}
  \label{eq:norad2}
k_{ab}=2\left(D_aD_b+h_{ab}\right)\Phi\,, \qquad \left(D^2+3\right)\Phi=0\,.
\end{equation}
In the absence of gravitational radiation, $k_{ab}$ therefore admits the decomposition \eqref{eq:norad2}. From \eqref{eq:STalpha}, we conclude that the scalar potential $\Phi(x^a)$ transforms by a simple shift under supertranslations,
\begin{equation}
  \label{eq:STGold}
\delta_\omega \Phi=\omega\,.
\end{equation}
Hence, $\Phi(x^a)$ is the Goldstone mode associated with spontaneous breaking of supertranslations, and will be referred to as the \emph{supertranslation mode}.

\section{Matching to null infinity}
\label{sec:scrimatch}

The relation between supertranslations at spatial infinity $i^0$, which we discussed in the previous section, and supertranslations at both past and future null infinity $\scri^\pm$ was analyzed at the linearized level in \cite{Troessaert:2017jcm} and at the nonlinear level in \cite{Prabhu:2019fsp}; see also \cite{Prabhu:2019daz}. We use this to clarify the relation of the Goldstone mode $\Phi$ of broken spi-supertranslations to its counterpart at null infinity. The latter is usually defined starting from the metric in Bondi coordinates around $\scri^+$,
\begin{align}
\label{eq:Bondi}
\dd s^2=&-\dd u^2-2 \dd u \dd r+r^2 \gamma_{AB}\dd x^A\dd x^B\\
\nonumber
&+\frac{2m}{r}\dd u^2+r\, C_{AB}\dd x^A \dd x^B+D^AC_{AB}\dd u \dd x^B+...\,,
\end{align}
where $x^A$ denote coordinates on the sphere and $\gamma_{AB}$ is the usual round metric with associated covariant derivative $D_A$. Asymptotic flatness at null infinity imposes restrictions on the admissible initial data. Sufficient conditions were derived by Christodoulou and Klainerman \cite{Christodoulou:1993uv}, which   imply that close to $\scri^+_-$ the shear $C_{AB}$ must be of the form \cite{Strominger:2013jfa,Compere:2018aar}
\begin{equation}
\label{eq:scriC}
C_{AB}|_{\scri^+_-}=-2D_AD_BC+\gamma_{AB}D^2 C\,,
\end{equation}
for an arbitrary function $C(x^A)$ on the sphere. Under a supertranslation generated by the vector field $\xi=T(x^A) \partial_u+...$, this function transforms as
\begin{equation}
  \label{eq:Ctrans}
\delta_{T} C=T\,,
\end{equation}
which shows that it is the Goldstone mode of broken supertranslations at future null infinity; for more details see \cite{Strominger:2017zoo}.

The relation between $C(x^A)$ and $\Phi(x^a)$ will follow from that between the supertranslation symmetry parameters $T(x^A)$ and $\omega(x^a)$, that has been described in detail by Troessaert \cite{Troessaert:2017jcm}. To make this precise, we cover the hyperboloid $\mathcal{H}$ with global coordinates,
\begin{equation}
ds^2=-\dd \tau^2+\cosh^2 \tau\, \gamma_{AB}\dd x^A\dd x^B\,,
\end{equation}
which we use to solve the equation \eqref{eq:STmode} that $\omega(x^a)$ satisfies. Defining $s\equiv \tanh \tau$ and decomposing over spherical harmonics,
\begin{equation}
\omega_l(s,x^A)=y_l(s)Y_{l,m}(x^A)\,, \qquad Y_{l,m}(-x^A)=(-1)^lY_{l,m}(x^A)\,,
\end{equation}
the wave equation \eqref{eq:STmode} gives rise to an ordinary differential equation,
\begin{equation}
  \label{eq:ODE}
(1-s^2)y''_l+\left(l(l+1)-\frac{3}{1-s^2}\right)y_l=0\,.
\end{equation}
Its solutions are spanned by\footnote{Let us mention for completeness that this parametrization in terms of Legendre functions actually misses the four lowest parity even modes. Although they can be straightforwardly constructed, we do not need their explicit expressions.}
\begin{equation}
(1-s^2)^{1/2}P^{2}_l(s)\,, \qquad  (1-s^2)^{1/2}Q^{2}_l(s)\,,
\end{equation}
where $P^{2}_l\,, Q^{2}_l$ are associated Legendre functions `on the cut' \cite{Magnus:1966formulas} that obey the parity properties
\begin{equation}
P^{2}_l(-s)=(-1)^l P^{2}_l(s)\,, \qquad Q^2_l(-s)=(-1)^{l+1} Q^2_l(s)\,.
\end{equation}
Hence, $\omega$ takes the general form
\begin{equation}
  \label{eq:omegasol}
  \omega=\omega^{\textrm{even}}+\omega^{\textrm{odd}}\,,
\end{equation}
with
\begin{align}
\omega^{\textrm{even}}(s,x^A)&=\sqrt{1-s^2}\sum_{l,m}\omega^{\textrm{even}}_{l,m} P^{2}_l(s)Y_{l,m}(x^A)\,,\\
\omega^{\textrm{odd}}(s,x^A)&=\sqrt{1-s^2}\sum_{l,m}\omega^{\textrm{odd}}_{l,m}Q^{2}_l(s)Y_{l,m}(x^A)\,.
\end{align}
These contributions are even and odd, respectively, under the antipodal map $(s,x^A) \mapsto (-s,-x^A)$ on the hyperboloid $\mathcal{H}$. In the limit $s \to 1$ corresponding to the future boundary of the de Sitter hyperboloid, the asymptotic behavior of the functions is given by
\begin{align}
  \label{eq:omegaexp}
  \omega^{\textrm{odd}}&\sim(1-s)^{-1/2}\sum_{l,m}\omega^{\textrm{odd}}_{l,m} Y_{l,m}(x^A)+O((1-s)^{1/2})\,,\\
  \omega^{\textrm{even}}&\sim (1-s)^{3/2}\sum_{l,m}\omega^{\textrm{even}}_{l,m}Y_{l,m}(x^A)+O((1-s)^{5/2})\,.
\end{align}
A similar expansion holds around $s=-1$. Note that the above discussion equally applies to the functions $\omega\,, \sigma$ or $\Phi$ since they all satisfy the same massive scalar equation.

Whether metrics of the form \eqref{eq:BS} evolve to asymptotically flat spacetimes at null infinity is a highly non-trivial question which has not been answered in full generality. It was nevertheless shown that regularity of the Weyl tensor at $\scri$ restricts the function $\sigma$ appearing in \eqref{eq:BS} to be parity-even \cite{Herberthson:1992gcz,Troessaert:2017jcm,Prabhu:2019fsp},
\begin{equation}
  \label{eq:sigmaodd}
\sigma(-s,-x^A)=\sigma(s,x^A)\,.
\end{equation}
This condition was suggested previously in \cite{Ashtekar:1990gc}, and similar parity conditions are also commonly used in the Hamiltonian framework \cite{Regge:1974zd,Henneaux:2018hdj}. Also note that this condition implies the falloff behavior $\sigma \sim (1-s)^{3/2}$, which can be used to show that the term proportional to $\delta \sigma$ in \eqref{eq:delta S log} actually vanishes in the limit where the Cauchy surfaces $\Sigma_\pm$ are pushed to the far past and future. In such case, the counterterm $S^{(\sigma)}$ is not needed for a well-posed action principle.   

Assuming \eqref{eq:sigmaodd} and evaluating the supertranslation charges \eqref{eq:STcharges} on the surface $s=0$, one then finds that it vanishes for parity-even supertranslations. These are therefore considered to generate proper gauge transformations on the phase space at spatial infinity. According to \cite{Troessaert:2017jcm,Prabhu:2019fsp} one can further relate the symmetry parameter $T(x^A)$ of supertranslations at future null infinity to the symmetry parameter $\omega(s,x^A)$ at spatial infinity via
\begin{equation}
\label{eq:matching}
T(x^A)=\lim_{s\rightarrow 1}\sqrt{1-s^2}\, \omega(s,x^A)=\lim_{s\rightarrow 1}\sqrt{1-s^2}\, \omega^{\textrm{odd}}(s,x^A)\,.
\end{equation}
Since $T(x^A)$ is expressed in terms of the parity-odd function $\omega^{\text{odd}}$, this directly yields the antipodal matching condition advocated by Strominger \cite{Strominger:2013jfa}.

Having matched the supertranslation parameters, the respective transformation properties \eqref{eq:STGold} and \eqref{eq:Ctrans} imply that the Goldstone modes $C$ and $\Phi$ are similarly related by
\begin{equation}
  \label{eq:Goldstonematch}
C(x^A)=\lim_{s\to 1}\sqrt{1-s^2}\, \Phi(s,x^A)\,.
\end{equation}
We will make further use of this relation in the next section.

\section{Infrared effective action on the celestial sphere}
\label{sec:effective action}
We finally come to the main motivation of this work, namely the derivation of the celestial two-point function \eqref{eq:C correlator} and its explicit relation to gravitational IR divergences. We reviewed in section~\ref{sec:inought} the fact that the variation of the Einstein--Hilbert action contains logarithmic IR divergences, which are naturally inherited by the symplectic structure of the theory. The Comp\`ere--Dehouck counterterm $\SCD$ given in \eqref{eq:S extra} precisely cures these divergences in a way that yields a finite symplectic structure \cite{Compere:2011ve}.\footnote{The logarithmic divergence in the symplectic structure stems from the logarithmically divergent term $\Delta\mathcal{R}^{(log)}_\mathcal{S}$ which is cancelled by a contribution from $\SCD$. The other divergent terms in \eqref{eq:S total} do not contribute to the symplectic structure.} Similarly, the total action \eqref{eq:S total} diverges logarithmically on-shell if $\SCD$ is not included. In other words, $\SCD$ fully controls the IR-divergent sector of the theory that is the object of interest here. In particular, one might guess that it controls the supertranslation Goldstone mode and constrains its two-point function. We show below that this is indeed the case.

We evaluate the Comp\`ere--Dehouck action \eqref{eq:S extra} on field configurations with vanishing magnetic Weyl tensor. For such radiation-free configurations, $k_{ab}$ can be simply written in terms of the spi-supertranslation mode $\Phi$ through \eqref{eq:norad2}. We find that the action consequently localizes at the corners $\mathcal{S}_\pm$ which are identified with the celestial sphere in the limit $\Lambda \to \infty$. Only a subset of $\Phi$ configurations turn out to be extrema of this action, in a way that \textit{explicitly breaks supertranslation symmetry} down to global translations. Furthermore, the constraint satisfied by extremum configurations takes the form of a differential equation whose Green's function precisely coincides with the celestial correlator \eqref{eq:C correlator}.

It is convenient to expand $k_{ab}$ and $\sigma$ in a Fefferman-Graham expansion and express the onshell action in terms of their two asympotically independent modes. For this, we first write the de Sitter metric $h_{ab}$ in planar coordinates,
\begin{equation}
ds_{\mathcal{H}}^2=\eta^{-2}\left(-\dd \eta^2+\delta_{ij} \dd x^i \dd x^j\right)\,, \qquad i,j=1,2\,.
\end{equation}
Planar coordinates only cover half of the hyperboloid, with the future conformal boundary $\mathcal{S}_+$ lying at $\eta=0$ and a cosmological horizon lying at $\eta \to \infty$. In these coordinates the geometry of the boundary $\eta=0$ appears as that of the euclidean plane, and may be completed by the point at infinity in order to recover the sphere as the conformal boundary of the global de Sitter hyperboloid.

From \eqref{eq:k conditions} and \eqref{eq:eom sigma and k}, we see that both $\Phi$ and $\sigma$ satisfy the same Klein--Gordon equation with negative mass squared $m^2=-3$. Its solutions admit the following Fefferman--Graham expansion close to $\mathcal{S}_+$,
\begin{equation}
\label{eq:FG expansion}
\Phi(\eta,x)=\eta^{-1} \left( \Phi_{(0)}+\eta^{2}\, \Phi_{(2)}+\eta^4 \ln \eta\, \tilde{\Phi}+\eta^{4}\, \Phi_{(4)}+O(\eta^6) \right)\,,
\end{equation}
and similarly for $\sigma$, where all terms in this expansion can be expressed in terms of the two independent asymptotic modes $\Phi_{(0)}$ and $\Phi_{(4)}$.\footnote{\label{footnote:logarithm} The two independent asymptotic modes are actually related to one another through a \textit{nonlocal} relation involving a choice of de Sitter boundary-bulk propagator. In momentum space this relation is found to be $\Phi_{(4)}(k)\sim k^4 \log k\,\Phi_{(0)}(k)$, where the appearance of a logarithm is directly related to the regularization of the corresponding two-point function \cite{Bzowski:2015pba}.} 
For our purposes, it is enough to write down the expression for the subleading coefficient,
\begin{equation}
\label{eq:Phi_2}
\Phi_{(2)}=-\frac{1}{4} \square \Phi_{(0)}\,.
\end{equation}

Before proceeding further, it is instructive to work out the relation between the Goldstone mode of supertranslations at null infinity defined on the sphere $C(x^A)$ and its counterpart at spatial infinity defined on the plane $\Phi_{(0)}(x^i)$. For convenience, we cover the sphere with stereographic coordinates, i.e., we make the replacement $x^A \mapsto x^i_{\text{stereo}}$. The transformation between global $(\tau,x^i_{\text{stereo}})$ and planar $(\eta,x^i)$ coordinates is fairly complicated but in the limit $\tau \to \infty$ simply reduces to
\begin{equation}
\label{eq:global to planar}
\eta \sim e^{-\tau} (1+|x|^2)\,, \qquad x^i \sim x^i_{\text{stereo}}\,.
\end{equation}
Together with \eqref{eq:Goldstonematch}, we thus find
\begin{equation}
\label{eq:C Phi0}
C(x)=\frac{2\, \Phi_{(0)}(x)}{ (1+|x|^2)}\,,
\end{equation}
which can be recognized as the transformation of a conformal primary of weight $\Delta=-1$ from the sphere to the plane.\footnote{A field $O_\Delta$ that is a coordinate scalar and transforms under Weyl rescalings as
\begin{equation}
\nonumber
O_\Delta'=\Omega^{-\Delta} O_\Delta\,, \qquad g'_{ij}=\Omega^2 g_{ij}\,,
\end{equation}
defines a conformal primary field of scaling dimension $\Delta$. Indeed, under the combined coordinate change $y^{i}=\lambda x^i$ and Weyl rescaling $g'_{ij}=\lambda^2 g_{ij}$, we have
\begin{equation}
\nonumber
O_{\Delta}'(y)=\lambda^{-\Delta} O_{\Delta}(x), \qquad g'_{ij}(y)=g_{ij}(x)\,,
\end{equation}
which is the usual transformation of a primary field under the corresponding conformal transformation.}
Indeed, the metric on the sphere is conformally related to that on the plane by
\begin{equation}
ds^2_{\text{sphere}}=\Omega(x)^2 ds^2_{\text{plane}}\,, \qquad \Omega(x)\equiv \frac{2}{1+|x|^2}\,,
\end{equation}
such that \eqref{eq:C Phi0} can be written
\begin{equation}
\label{eq:planesphere}
C(x)=\Omega(x)\, C_{\text{plane}}(x)\,, \qquad C_{\text{plane}}(x)\equiv\, \Phi_{(0)}(x)\,.
\end{equation}
The leading term in the Fefferman--Graham expansion \eqref{eq:FG expansion} is therefore identified with the corresponding celestial conformal field on the plane.

Using \eqref{eq:norad2} together with the Fefferman-Graham expansion \eqref{eq:FG expansion}, the Comp\`ere-Dehouck action evaluates to
\begin{align}
\label{eq:S onshell}
S^{(k)}&=-\frac{1}{4} \int d^2x \left( 8\Phi_{(2)}^2+\left(\square \Phi_{(0)} \right)^2+4 \Phi_{(2)} \square \Phi_{(0)} \right)=-\frac{1}{8} \int d^2x \left(\square \Phi_{(0)} \right)^2\,,\\
S^{(\sigma)}&=\frac{1}{2} \int d^2x\, \left(-\eta^{-4} \sigma_{(0)}^2+(1+2\log \eta) \sigma_{(0)} \tilde{\sigma}+\sigma_{(2)}^2+2\sigma_{(0)}\sigma_{(4)} \right)=0\,.
\end{align}
The vanishing of $S^{(\sigma)}$ occurs due to the parity condition \eqref{eq:sigmaodd}. Indeed, the expansion \eqref{eq:omegaexp} applied to $\sigma$ together with the change of coordinate \eqref{eq:global to planar} implies  $\sigma_{(0)}=\sigma_{(2)}=\tilde{\sigma}=0$. We thus arrive at the \emph{infrared effective action}
\begin{equation}
\label{eq:celestials}
S_{\text{IR}}\equiv \SCD\big|_{\text{onshell}}=-\frac{\log \Lambda}{32\pi G} \int d^2x \left(\square C_{\text{plane}} \right)^2\,.
\end{equation}

The first important observation is that extrema of this action satisfy the constraint
\begin{equation}
\label{eq:box2}
\square^2 C_{\text{plane}}=0\,.
\end{equation}
Taking the flat metric to be spanned by complex coordinates, $\dd s^2=\dd z \dd \bar{z}$, the explicit solutions to this equation read
\begin{equation}
    C_{\text{plane}}=c_1(z)+z c_2(\bar z)+c_3(\bar z)+\bar{z} c_4(z).
\end{equation}
When mapped back to the sphere through \eqref{eq:planesphere}, the only real and regular solutions are the four lowest spherical harmonics spanned by $1, z, \bar{z}, z \bar{z}$, which correspond to \textit{global translation} configurations. Thus, the infrared effective action \textit{explicitly breaks supertranslation symmetry} down to the global subgroup of translations.

Finally, we recover the celestial two-point function \eqref{eq:C correlator} up to normalization as the Green's function associated with \eqref{eq:box2}, 
\begin{equation}
\label{eq:correlator}
\langle C_{\text{plane}}(x)C_{\text{plane}}(0)\rangle\sim G\, |x|^2\log |x|^2\,,
\end{equation}
where the appearance of Newton's constant follows from dimensional analysis. Alternatively, in the original gravitational action \eqref{eq:S extra} it can be seen that it is the `graviton' field $\tilde{k}_{ab}=G^{-1/2} k_{ab}$ which has canonical kinetic term rather than $k_{ab}$ itself. Note that it had been previously realized that the action \eqref{eq:celestials} is the one required to recover the celestial two-point function \eqref{eq:correlator} \cite{Kalyanapuram:2020epb,Kalyanapuram:2021bvf}. In this work we showed how it arises from a careful study of asymptotically flat gravity near spatial infinity. 

The celestial two-point function \eqref{eq:correlator} also appears in the description of the displacement memory effect.\footnote{For a review of the displacement memory effect, see \cite{Strominger:2014pwa,Strominger:2017zoo, Compere:2018aar} and references therein.} We can actually compare the constraint \eqref{eq:box2} derived from the the infrared effective action to the equation satisfied by the total supertranslation shift $\Delta C$ over a (retarded) time interval $\left[u_i,u_f \right]$ \cite{Compere:2018aar}, 
\begin{equation}
\label{eq:memory}
-\frac{1}{4}(D^2+2)D^2 \Delta C=\Delta m+\int_{u_i}^{u_f} du\, T_{uu}\,,
\end{equation}
where
\begin{equation}
T_{uu}=\frac{1}{8} N_{AB}N^{AB}+4\pi G \lim_{r \to \infty} r^2 T^M_{uu}\,.
\end{equation}
This shift $\Delta C(x)$ is sourced by the total loss of Bondi mass aspect $\Delta m_B(x)<0$, the flux of gravitational radiation carried by the News tensor $N_{AB}$, and the flux of massless matter with stress tensor $T^M$. Equation \eqref{eq:memory} involves quantities defined on the sphere rather than the euclidean plane. Using \eqref{eq:planesphere} and $D^2=\Omega^{-2}\square$, we can rewrite the left-hand side of \eqref{eq:memory} in terms of quantities on the plane,
\begin{equation}
\label{eq:D4 operator}
(D^2+2)D^2 \Delta C=(\Omega^{-2}\square +2)\Omega^{-2} \square ( \Omega \Delta C_{\text{plane}})=\Omega^{-3} \square^2 \Delta C_{\text{plane}}\,.
\end{equation}
Hence, the constraint \eqref{eq:box2} implied by the infrared effective action \eqref{eq:celestials} is nothing but the memory equation \eqref{eq:memory} pulled back to $\scri^+_-$ where source terms on the right-hand side are absent. It also follows that the celestial two-point function \eqref{eq:correlator} acts as euclidean propagator for the memory equation \eqref{eq:memory}, a point already made explicit in \cite{Strominger:2017zoo}.

To close this section, we note that the infrared effective action \eqref{eq:celestials} is invariant under $\text{SL}(2,\mathbb{C}) \approx \text{SO}(3,1)$ global conformal transformations, which is precisely the action of the four-dimensional Lorentz group on the celestial sphere. However, it is not invariant under local conformal transformations in contrast to more  conventional two-dimensional CFTs. In the extended BMS group of asymptotic symmetries at null infinity \cite{Barnich:2010eb}, such local conformal transformations are generated on the celestial sphere by Virasoro superrotations, but their analogue at spatial infinity has not been studied so far. Even if these could be incorporated into the present analysis, in its present form the infrared effective action would break the extended BMS group down to the Poincar\'e group.

\section{Path integral representation of the soft factor}
\label{sec:path integral}
We end by showing that the the infrared effective action \eqref{eq:celestials} finds a natural place in a path integral formulation of the celestial CFT, and can be employed to re-derive the IR-divergent soft factor $\mathcal{A}_{\text{soft}}$ \eqref{eq:A soft} when used in the context of scattering amplitudes. Our starting point for this derivation is the formulation of the soft factor as the expectation value\footnote{Similar representations of the soft factor as the expectation value of \textit{gravitational Wilson lines} have been discussed in \cite{Naculich:2011ry,White:2011yy,Melville:2013qca}.} \cite{Himwich:2020rro,Arkani-Hamed:2020gyp}
\begin{equation}
\label{eq:expectation}
\mathcal{A}_{\text{soft}}(p_1,..,p_n)= \langle \mathcal{W}_i\, ...\, \mathcal{W}_n \rangle\,,
\end{equation}
where there is one vertex operator insertion per hard external particle,
\begin{equation}
\label{eq:vertex}
\mathcal{W}_i=e^{i \eta_i \omega_i \Cp(x_i) }\,.
\end{equation}
This formulation was used in \cite{Himwich:2020rro} to infer the two-point function \eqref{eq:C correlator} of the supertranslation Goldstone mode. Here, we propose instead to directly evaluate the soft factor \eqref{eq:expectation} as a path integral weighted by the infrared effective action \eqref{eq:celestials}, 
\begin{equation}
\label{eq:path integral}
\mathcal{A}_{\text{soft}}(p_1,..,p_n)=\langle \mathcal{W}_i\, ...\, \mathcal{W}_n \rangle=\int \mathcal{D}\Cp\, \mathcal{W}_i\, ...\, \mathcal{W}_n \, e^{-S_{\text{IR}}[\Cp]/\log^2 \Lambda} \,.
\end{equation}
The $\log^2\Lambda$ factor in the exponent has been chosen such as to yield the normalization of the two-point function given in \eqref{eq:C correlator}.
This is a Gaussian integral that we can straightforwardly perform, yielding
\begin{equation}
\label{eq:A soft final}
\mathcal{A}_{\text{soft}}(p_1,..,p_n) \sim \exp\left[- \frac{2G \log \Lambda}{\pi} \sum_{i,j=1}^n \eta_i \eta_j \omega_i \omega_j\,  |x_i-x_j|^2 \ln |x_i-x_j|^2 \right]\,,
\end{equation}
where an overall numerical factor in the exponent has been reabsorbed into $\log \Lambda$. We have recovered the standard result \eqref{eq:A soft} without having to assume the form of the two-point function of the supertranslation Goldstone mode. 

\section{Discussion}
\label{sec:discussion}

In this paper, we have shown that an \textit{infrared effective action} governing the celestial two-point function of the supertranslation Goldstone mode can be derived from the Comp\`ere--Dehouck action. The latter is used to render the symplectic structure and on-shell value of asymptotically flat gravity finite, and therefore controls the IR-divergent sector of the theory. As demonstrated in the last section, this effective action finds its natural place in a path integral formulation of the celestial CFT.

We should stress that our analysis crucially relies on isolating the Comp\`ere--Dehouck contribution from the total action \eqref{eq:S total}.\footnote{Alternatively, the same results could have been obtained by restricting our attention to the other part of the action, i.e.,  $S_{\text{total}}-\SCD$.} Indeed, when one evaluates this total action on radiation-free solutions as was done in section~\ref{sec:effective action}, one finds that the IR-divergent contribution \eqref{eq:S onshell} of interest precisely cancels against the other logarithmically divergent term $\log \Lambda\, (\mathcal{R}^{(log)}_{\mathcal{S}_+}-\mathcal{R}^{(log)}_{\mathcal{S}_-})$ in \eqref{eq:S total}. We explicitly demonstrate this in appendix~\ref{sec:GHY and MM}. This is as it should be since the total action \eqref{eq:S total} is precisely constructed such as to yield an IR-finite symplectic structure. Because our interest lies in the IR-divergent soft factors \eqref{eq:A soft} and their relation to the celestial two-point function \eqref{eq:C correlator}, we deliberately chose not to consider the fully regulated and IR-finite theory, and instead restrict our attention to the counterterm $\SCD$ that precisely encodes these divergences in the unregulated theory. This should also remind us that divergent scattering amplitudes such as \eqref{eq:factorization} are unobservable and therefore unphysical. A different way to set up the computation of observable quantities which would not yield divergent scattering amplitudes even at intermediate steps, such as that based of Faddeev--Kulish dressings \cite{Kulish:1970ut,Ware:2013zja,Kapec:2017tkm,Arkani-Hamed:2020gyp}, would appear more in line with the semiclassical treatment of gravity proposed by Comp\`ere and Dehouck based on the fully regulated action functional $S_{\text{total}}$ or a Hamiltonian approach proposed in \cite{Henneaux:2018hdj,Henneaux:2018cst} with finite action due to parity conditions.

We have seen in section~\ref{sec:effective action} that the discussion of the soft sector of four-dimensional asymptotically flat gravity reduces to that of fields on three-dimensional de Sitter spacetime. This makes it is very tempting to speculate that information about a dual celestial CFT can be obtained by applying a form of dS/CFT correspondence along the lines of \cite{deBoer:2003vf,Ball:2019atb}, since the celestial sphere precisely plays the role of the de Sitter conformal boundary. In the case at hand, the supertranslation Goldstone mode appears as one of the two asymptotically independent modes in the Fefferman-Graham expansion \eqref{eq:FG expansion}. Within a dS/CFT correspondence, this mode would be identified with the vacuum expectation value of a CFT operator of scaling dimension \cite{Strominger:2001pn}\footnote{Note that a negative scaling dimension such as \eqref{eq:scaling} can only occur in a non-unitary CFT. Within AdS/CFT, this would be discarded on the basis that the corresponding bulk mode $\Phi_{(0)}$ is not normalizable under the Klein--Gordon inner product and therefore cannot be allowed to fluctuate \cite{Breitenlohner:1982bm,Breitenlohner:1982jf}. Such a restriction does not occur in de Sitter space since both $\Phi_{(0)}$ and $\Phi_{(4)}$ are normalizable.}
\begin{equation}
\label{eq:scaling}
\Delta=1- \sqrt{1-m^2}=-1\,, \qquad m^2=-3\,,
\end{equation}
which simply coincides with the order at which this mode enters the Fefferman-Graham expansion~\eqref{eq:FG expansion}. This is indeed the correct scaling dimension of the supertranslation Goldstone mode \cite{Himwich:2020rro,Pasterski:2021fjn}. In this case, the mode $\Phi_{(4)}$ with $\Delta=3$ would acquire the interpretation of a source for $\Phi_{(0)}$. It seems that this set-up would appropriately capture the \emph{Goldstone diamond} for supertranslations introduced in \cite{Pasterski:2021fjn,Pasterski:2021dqe}, and we find that the precise relation between $\Phi_{(0)}$ and $\Phi_{(4)}$ in momentum space involves a logarithm (see footnote~\ref{footnote:logarithm}). We leave the exploration of such an application of the dS/CFT correspondence to future studies.

In addition to supertranslations, soft theorems and memory effects have been recently related to additional symmetries of asymptotically flat gravity at null infinity such as superrotations \cite{Barnich:2010eb,Cachazo:2014fwa,Campiglia:2014yka,Pasterski:2015tva,Compere:2018ylh}, dual supertranslations \cite{Godazgar:2018qpq,Godazgar:2019dkh} or logarithmic supertranslations \cite{Laddha:2018myi}. Analogous symmetries can be realized in a Beig--Schmidt coordinate system like \eqref{eq:BS} near spatial infinity, although their precise relations to their cousins at null infinity has not been worked out. Thus, it would be very interesting to see whether one can construct a well-defined variational principle allowing for these symmetries and derive effective actions and correlation functions for the corresponding Goldstone modes, as was done in this work for supertranslations. Superrotations likely require one to promote the de Sitter metric $h_{ab}$ to a dynamical field, which would in turn strengthen the case for a form of the dS/CFT correspondence mentioned above.

\acknowledgments{We thank Mina Himwich and Andy Strominger for valuable comments. The work of KN is supported by a grant from the Science and Technology Facilities Council (STFC). The work of JS is supported by
the Erwin-Schr\"odinger fellowship J-4135 of the Austrian Science Fund
(FWF) and by the DOE grant de-sc/0007870.}

\appendix
\section{Variational principle and counterterms}
\label{app:variational principle}
In this appendix we provide details on the variational principle of the action \eqref{eq:totaction} following~\cite{Compere:2011ve}. We will make use of Stoke's theorem in the form
\begin{equation}
  \label{eq:Stoke}
  \int_{\mathcal M} \dd^4x\sqrt{-g}\, \nabla_\alpha v^{\alpha}=\int_{\mathcal M}\dd^4x\, \partial_\alpha(\sqrt{-g} v^{\alpha})=\int_{\partial \mathcal{M}}(\dd^3x)_\alpha \sqrt{-g}v^{\alpha}\,,
\end{equation}
and the conventions of \cite{Compere:2011ve},
\begin{equation}
  \label{eq:6}
  (\dd^{n-p}x)_{\mu_1...\mu_p}\equiv\frac{1}{p!(n-p)!}\epsilon_{\mu_1...\mu_n}\dd x^{\mu_{p+1}}\wedge...\wedge \dd x^{\mu_n}\,,
\end{equation}
with $(\dd^3x)_\tau=-\dd \rho\, (\dd^2x)_\tau,\,(\dd^3x)_\rho=-\dd \tau\, (\dd^2x)_\rho$, and $(\dd^2x)_\tau=-(\dd^2x)_\rho=-\dd^2S$ with $\dd^2S=\frac{1}{2}\epsilon_{AB}\dd x^A\wedge \dd x^B$ in a Beig--Schmidt coordinate system $(\rho,\tau,x^A)$.

Setting $16\pi G=1$ for convenience, variation of the Einstein--Hilbert action \eqref{eq:EH} produces boundary terms at the Cauchy surfaces $\Sigma_\pm$,
\begin{equation}
\delta S_{\text{EH}}|_{\Sigma_{\pm}}=\pm \int_{\Sigma_\pm}(\dd^3x)_\tau \Theta^\tau\,,
\end{equation}
where
\begin{equation}
\label{eq:Theta}
    \Theta^{a}(\dd^3x)_a=\sqrt{-g}\, (g^{a\lambda}\nabla^\nu\delta g_{\lambda\nu}-g^{\lambda \rho}\nabla^{a}\delta g_{\lambda \rho})(\dd^3x)_a\,.
\end{equation}
We will restrict our attention to the future Cauchy surface in what follows. Explicit evaluation of the above expression yields
\begin{gather}
  \delta S_{\text{EH}}|_{\Sigma_+}=-\log \Lambda_+ \int \dd^2S \sqrt{-h}\left(\frac{1}{2}\delta k^{bc}D^\tau k_{bc}+4 \delta \sigma D^\tau \sigma-\delta k^{bc}D_c k\indices{^\tau_b}\right)\nonumber\\
  +2\Lambda_+ \int \dd^2S \sqrt{-h}D^\tau\delta \sigma+\log^2 \Lambda_+ \,\delta\int \dd^2S \sqrt{-h} (D_b  i^{\tau b}-D^\tau  i)     \label{eq:2}
  \\
-\log \Lambda_+\, \delta \int \dd^2S \sqrt{-h} \left(-k^{bc}D^\tau k_{bc}+k^{bc}D_c k\indices{^\tau_b}-D_b  h^{(2)\tau b}+D^\tau  h^{(2)}- 10 \sigma D^\tau\sigma\right)+ O(\Lambda^0)\,, \nonumber
\end{gather}
after doing the trivial $\rho$-integral and pulling out total variations.
 The terms in the second and third line, which we will denote by $\Lambda_+\, \mathcal{R}_{\mathcal{S}_+}+\log \Lambda_+\, \mathcal{R}_{\mathcal{S}_+}^{(log)}+\log^2 \Lambda_+\mathcal{R}_{\mathcal{S}_+}^{(log\, 2)}$, are total variations that should be subtracted from the action. One should similarly find additional boundary terms and/or impose boundary conditions on the variations at $\Sigma_{\pm}$ such that the $ O(\Lambda^0)$ term in \eqref{eq:2} vanishes. This requires a careful analysis of the variational problem that lies outside of the scope of the present paper.

The term in the first line on the other hand is not a total variation and one must find another way to cancel it while still allowing for supertranslation. As demonstrated by Comp\`ere and Dehouck, the addition of $S_{\text{CD}}$ to the Einstein--Hilbert action precisely cancels the first line of \eqref{eq:2} in the variation of the action. Together with the boundary terms necessary to cancel the total variations $\mathcal{R}, \mathcal{R}^{(log)},\mathcal{R}^{(log\, 2)}$ and the $O(\Lambda^0)$ term in \eqref{eq:2}, this guarantees a well-defined action principle at $\Sigma_\pm$.

It remains to discuss the variational principle on the hyperbolic cutoff surface $\rho=\Lambda_+$. In order to have Dirichlet boundary conditions on the induced metric, one has to add the Gibbons--Hawking--York boundary term. However, this term is not enough to guarantee stationarity of the action principle for all variations compatible with the boundary conditions when $\Lambda$ is sent to infinity. This issue is resolved by adding the Mann--Marolf counterterm \cite{Mann:2005yr}, as discussed in the main text. For a demonstration of this fact in the case $k_{ab}\neq 0$ we refer to \cite{Virmani:2011gh}.

In summary, we have argued that the total action \eqref{eq:S total} yields a well-defined variational principle both on the the hyperbolic cut-off $\mathcal{H}_\pm$ and Cauchy surfaces $\Sigma_\pm$, up to the above-mentioned caveat concerning the  $O(\Lambda^0)$ terms in \eqref{eq:2}.

\section{Evaluation of the on-shell action}
  \label{sec:GHY and MM}

In this appendix we provide more details on the on-shell value of the total action \eqref{eq:S total}. In order to evaluate these terms we will need the expansion of Einstein's equations in the Beig--Schmidt system \eqref{eq:BS} up to second order. To first order, the equations of motion are solved by condition \eqref{eq:R condition} for $h_{ab}$, the gauge conditions \eqref{eq:k conditions} and the equations of motion for $\sigma$ and $k_{ab}$ \eqref{eq:eom sigma and k}. Next we find the equations for the logarithmic term
  \begin{align}
    \label{eq:9}
 i\equiv h^{ab}i_{ab}=D^ai_{ab}=0,\qquad (D^2-2)i_{ab}=0\,.
  \end{align}
  We are only going to make use of the second order equations
  \begin{subequations}
    \begin{align}
      \label{eq:10}
      h^{(2)}&\equiv h^{ab}h^{(2)}_{ab}=\frac{1}{4}k_{ab}k^{ab}+k_{ab}D^aD^b\sigma+12\sigma^2+D^a\sigma D_a\sigma\,,\\
      D^ch^{(2)}_{ca}&=\frac{1}{2}D^{c}k_{ab}k\indices{^b_c}+D_a\left(D_b\sigma D^b\sigma+8\sigma^2-\frac{1}{8}k^{bc}k_{bc}+k_{bc}D^bD^c\sigma\right)\,.
    \end{align}
  \end{subequations}

Starting with the action $S=S_{\text{EH}}+S_{\mathcal{H}_+}$ defined in \eqref{eq:totaction}, we note that the Einstein--Hilbert term vanishes identically when Einstein's equations are satisfied, and the only non-zero contribution comes from $S_{\mathcal{H}_+}$. Assuming that the Mann--Marolf tensor $\hat{K}_{ab}$ admits an asymptotic expansion of the form of
  \begin{equation}
    \label{eq:khatexp}
\hat{K}_{ab}=\rho\, \hat{\kappa}^{(0)}_{ab}+\hat{\kappa}^{(1)}_{ab}+\rho^{-1} \log \rho\, \hat{\kappa}^{(log)}_{ab}+\rho^{-1}\, \hat{\kappa}^{(2)}_{ab}+o(\rho^{-1})\,,
\end{equation}
the respective coefficients $\hat{\kappa}^{(0)}, \hat{\kappa}^{(1)},...$ can be determined recursively from the defining equation \eqref{eq:khatdef}. In particular, to leading order one finds  $\hat{\kappa}^{(0)}_{ab}=h_{ab}$ upon using \eqref{eq:R condition}. The solution to the subleading order equations can be found in \cite{Mann:2006bd} albeit without the logarithmic term. We won't display the rather lengthy expressions for $\hat{K}_{ab}$, and restrict our attention to its trace $\hat{K}=H^{ab}\hat{K}_{ab}$ which is the quantity required to evaluate the Mann--Marolf counterterm. We find
\begin{equation}
    \label{eq:kexpansion}
    \hat{K}=3\rho^{-1}+\rho^{-2} \hat{K}^{(1)}+\rho^{-3}\log \rho\, \hat{K}^{(log)}+\rho^{-3} \hat{K}^{(2)}+o(\rho^{-3})\,,
  \end{equation}
where the first two orders,
  \begin{align}
  \hat{K}^{(1)}&=D^2 \sigma +3\sigma +\frac{1}{4} D^a D^b k_{ab}\,,\\
  \hat{K}^{(log)}&=-\frac{1}{2} i-\frac{1}{4}D^2 i+\frac{1}{4}D^a D^b i_{ab}\,,
  \end{align}
vanish identically by means of the above equations of motion.
  The non-vanishing term $\hat{K}^{(2)}$ is given by
  \begin{align}
  \hat{K}^{(2)}=\frac{1}{4}\Big(&k^{ab}k_{ab}-27\sigma^2-4D_a \sigma D^a \sigma-5 k^{ab} D_aD_b \sigma+D_a D_b \sigma D^aD^b \sigma\Big)\,.
  \end{align}
  With the expansion for the extrinsic curvature of the induced metric on a constant $\rho$-slice
  \begin{equation}
    \label{eq:extk}
K=3 \rho^{-1}+\rho^{-3}\left(\frac{1}{4} k^{ab}k_{ab}-6 \sigma^2-D_a\sigma D^a\sigma-k^{ab}D_aD_b\sigma)+o(\rho^{-3}\right),
  \end{equation}
  we have for the combination $K-\hat{K}$ appearing in the boundary counterterm
  \begin{equation}
    \label{eq:tracekhat}
    K-\hat{K}=\frac{1}{4\rho^3}\left(3\sigma^2-D_aD_b\sigma D^aD^b\sigma+k^{ab}D_aD_b\sigma\right)+o(\rho^{-3})\,.
  \end{equation}
After integration over $\rho$, the on-shell contribution to $S$ is therefore of order $O(\Lambda_\pm^0)$.

Let us now turn our attention to the other terms in the total action $S_{\text{total}}$ \eqref{eq:S total}. 
We find
\begin{subequations}
  \begin{align}
    \label{eq:11}
    \mathcal{R}_{\mathcal{S}_+}&=2\int \dd^2S \sqrt{-h}\, D^\tau \sigma\,,\\
    \mathcal{R}^{(log)}_{\mathcal{S}_+}&=\frac{1}{4}\int\dd^2S\sqrt{-h} \left(2k_{ab}D^bk^{a\tau}-k_{ab}D^\tau k^{ab}-2\sigma D^\tau \sigma\right)\,,\\
    \mathcal{R}^{(log\, 2)}_{\mathcal{S}_+}&=0\,.
\end{align}
\end{subequations}
We recognize the term  $\mathcal{R}_{\mathcal{S}_+}$ as being proportional to the supertranslation charge \eqref{eq:STcharges} for $\omega=1$, which is nothing but the total mass. Since the supertranslation charges are independent of the slice chosen for evaluating them, this contribution will cancel against the term at $\mathcal{S}_-$. Finally, we see that the Comp\`ere--Dehouck counterterm \eqref{eq:S extra} yields the same on-shell value as $ \mathcal{R}^{(log)}_{\mathcal{S}_+}$. Since they appear with opposite signs in the total action, they cancel against each other as mentioned in the discussion. The full Einstein--Hilbert action therefore is $O(\Lambda^0)$, receiving contributions from the $O(\Lambda^0)$ counterterms on the Cauchy slices $\Sigma_\pm$ and the combination \eqref{eq:tracekhat} from the counterterms on the hyperbolic cut-off surfaces $\mathcal{H}_\pm$.

\bibliography{bibl}
\bibliographystyle{fullsort.bst}
\end{document}